# Direct observation of two-phonon bound states in ZnTe

Jianbo Hu[1,2], Oleg V. Misochko[3], and Kazutaka G. Nakamura[1,2,*]

[1]*Materials and Structures Laboratory, Tokyo Institute of Technology*

*R3-10, 4259 Nagatsuta, Yokohama 226-8503, Japan*

[2]*JST-CREST, Sanban-cho bldg, 5, Sanban-cho, Chiyoda, Tokyo 102-0075, Japan*

[3]*Institute of Solid State Physics, Russian Academy of Sciences*

*142432 Chernogolovka, Moscow region, Russia*

A coherent two-phonon bound state has been impulsively generated in ZnTe(110) via second-order Raman scattering in the time domain for the first time. The two-phonon bound state, composed of two anticorrelated in wave vector acoustic phonons, exhibits full $\Gamma_1$ symmetry and has energy higher than the corresponding 2TA(X) overtone. By suppressing two-phonon fluctuations with a double-pulse excitation, the coexistence of coherently excited bound and unbound two-phonon states has been demonstrated.

# I. INTRODUCTION

In order to verify the possible occurrence of bound states in crystals, studies are currently in progress on the vibrational excitations in overtone and combination regions. Sharp "impurity like" energy levels may occur in these regions lying outside the multiphonon continua because of the residual anharmonic phonon-phonon interaction. In the simplest case, these levels are related to the two-phonon bound state (TPBS) [1]. Such a TPBS was first observed in diamond [2], and theoretically discussed two decades later [3]. Since then, a number of experimental and theoretical efforts [4-6] have been devoted to improve our understanding of this unique state, even though some controversies still exist. The TPBS in general cannot be separated into two free phonons without spending enough energy. Despite its composite nature, efficient decay routes for TPBS do not exist because the states are not embedded in the multiphonon continua. Thus, the direct dissociation of the TPBS into its component phonons does not play a role in its relaxation. Other relaxation mechanisms, either depopulation or dephasing, mediated through single-site anharmonicity, considerably restrict their efficiencies.

The TPBS has been extensively studied in the frequency-domain [1]. In contrast, the TPBS studies in the time-domain, where it is possible to create bound states in an off-equilibrium condition, have been few but already demonstrated promise in extracting additional important information [7,8]. Nonetheless, the prospect to excite bound phonons coherently has not been realized yet, and the theoretical prediction that the localization degree of the bound state can be manipulated [3] has not been experimentally demonstrated. In this paper, we report the first time-domain observation of the impulsively excited coherent TPBS in ZnTe by using femtosecond laser pulses.

ZnTe belongs to a class of electro-optic crystals having ultrafast response times which are of enormous technological and scientific importance. Its optical properties have been extensively studied in the frequency-domain. Based on Raman scattering [9,10] and infrared absorption spectra [11,12], the first- and second-order vibrational modes of ZnTe were accurately assigned. Some of these modes were also detected in the time-domain [13,14]. Recently, a comprehensive time-domain study on ZnTe(100) demonstrated that, besides the LO(Γ) mode at 6.15 THz, a mode centered around 3.5 THz can be impulsively excited using ultrashort laser pulses [13]. The assignment of this mode, however, remains controversial: both the TO(X)−TA(X) difference and the 2TA(X) overtone were suggested [12,13] to be responsible for its appearance. It should be also noted that one of the time-domain studies assigned the 3.5 THz feature to the LA(X) mode [14]. Such an unusual assignment was made to account for somewhat anomalous temperature dependence, resembling that of a single phonon. The ultimate assignment of the 3.5 THz mode is hindered by the fact that its intensity in both the frequency- and time-domains is significantly weaker than that of the first-order modes. However, the pump-polarization dependence of the LO(Γ) mode [13] in ZnTe suggests that it is possible to excite the second-order phonon mode alone by selecting an appropriate measurement geometry. In order to realize this possibility, we used in our study a ZnTe(110) single crystal. For this crystal orientation, the first-order Raman scattering is weak for the TO(Γ) mode due to resonance condition and prohibited for the LO(Γ) mode by symmetry [10]. Thus, only second-order phonon modes can be excited and observed using ultrashort laser pulses.

## II. EXPERIMENT

The sample was a $10\times10\times1$ mm$^3$ single crystal plate of ZnTe grown along the <110> axis by a seeded vapor-phase free growth method. Judging from the dislocation density of $5\times10^5$ cm$^{-2}$ the crystal quality was reasonably high. All measurements were made at room temperature. The method used is ultrafast pump-probe technique in which the stronger pump pulse drives the crystal into an excited time-varying state, which perturbs the weaker probe pulse that follows behind. The detected signal is the transmitted intensity of the probe beam as a function of the time delay $t$ of the probe relative to the pump pulse. More exactly we detect the differential transmission $\Delta T \equiv T - T_0$, that is the difference in transmission of the probe with $T$ and without $T_0$ the pump at variable probe delays using an isotropic detection scheme. A mode-locked Ti-sapphire laser provided pulses with the duration of ~40 fs centered at ~800 nm. The oscillator had a repetition rate of 80 MHz, providing an average power of 60 mW for the pump beam and 2 mW for the probe beam both of which were focused to a 50-μm-diameter spot. The polarizations of the pump and probe beams were orthogonal. For the two-pump pulse excitation, a stronger pump pulse was fed into a Michelson-type interferometer and split into two collinear pulses (of 23 and 30 mW average power) with a variable interpulse separation [15].

### III. RERULTS AND DISCUSSION

Figure 1 shows the normalized transient signal $\Delta T / T_0$ of the ZnTe(110) crystal excited using single pump pulse with an average power of 30 mW. Following a sharp change in transmission induced by the electronic response at $t = 0$, a small but well reproducible modulation exists due to impulsively excited lattice vibrations. Fitting the coherent part of the transient signal in the time-domain with a damped harmonic function allows us to determine the initial phase, lifetime, frequency and amplitude of the oscillation. The frequency and lifetime can be also obtained from a Fourier transform of the time domain signal shown in the lower inset of Fig. 1. The Fourier spectrum demonstrates only a single peak with a frequency of 3.67 THz, which is close to, but a bit higher than that of the 2TA(X) overtone in thermally excited ZnTe crystal [9].

To facilitate the assignment of the 3.67 THz lattice mode, we first carried out pump-polarization dependence measurement the results of which are shown in Fig. 2 (a). The results prove that the amplitude of the oscillatory signal is essentially independent of the pump polarization. They are in agreement with a completely symmetric entity: one of the representations contained in $X=\Gamma_1+2\Gamma_{12}+\Gamma_{15}$, which is appropriate for two transverse acoustic phonons from the X-point of the Brillouin zone. Thus, the symmetry consideration allows us to discard both the TO(X)−TA(X) difference and the one-phonon LA(X) mode as the source of coherent oscillations, since the both modes must demonstrate an angle dependence due to non-zero off-diagonal components of their Raman tensors. As compared to the 2TA(X) overtone observed in [9] at the same temperature, a small blue shift in frequency of 0.4 THz is indicative of a bound state of the two-phonon mode generated by a repulsive interaction [3]. To the best of our knowledge, it is the first observation of the two-acoustic phonon bound state theoretically predicted long ago [16]. Kamaraju et al. [14] reported coherent oscillations at 3.5 THz and assigned them as the LA(X) mode because of the observed defect-concentration and temperature dependences. However, the excitation power and crystals used in [14] are quite different from the present experiments. For instance, the excitation power was 3 orders magnitude higher than that of the present experiment, which resulted in photo-induced carriers created via a two-photon absorption process. Kamaraju et al. [14] also observed the defect-density dependence as strong phonon oscillation in a high defect-density sample but no coherent phonons in a high quality sample. In contrast, in our study the coherent phonons are observed in the high quality sample. The temperature dependence may also be affected by the densities of photo-induced carriers and defects, which were not accounted in their calculation. In addition, the observed polarization dependence is not consistent with the LA(X) phonon symmetry. Moreover, since the TPBS is quite similar to the single particle state bound to the impurity, the observed dependence on the crystal quality [14] can be explained without invoking the relaxation of wave vector conservation during the excitation process.

For the coherent one-phonon excitation, which changes the lattice polarizability, the atomic movement modulates inter-particle separation leaving the center of mass intact. However, in our case the change in transmission is induced by second-order Raman scattering $\Delta T \equiv T - T_0 = \sum_q \frac{\partial^2 T}{\partial Q_q \partial Q_{q'}} \langle Q_q Q_{q'} \rangle$, where $q$ is the wave vector, $Q$ is the normal mode operator and the average is over the phonon states [17]. Because of phase coherent excitation the average is reduced to $\langle Q_q Q_{-q} \rangle = \langle \Delta u^2(\pm q, t) \rangle$ with $u$ being the atomic displacement since the lattice is excited into two phonon modes characterized by equal frequencies and equal but opposite wave vectors. By virtue of the conservation principles that govern their creation, the two $|q_+\rangle$ and $|q_-\rangle$ phonons interfere destructively resulting in two-phonon coherence. We note in passing that the two phonon state generated by the second order Raman process can result in a squeezed lattice state [17, 18].

To gain further insight into the properties of the TPBS in ZnTe, we measured the dependence of the amplitude of coherent oscillations on the excitation intensity. As shown in Fig. 2(b), the pump-power dependence exhibits a linear dependence for the amplitude, suggesting the generation mechanism for TPBS most likely to be off-resonant impulsive stimulated second-order Raman scattering [19]. Indeed, the photon energy of 1.55 eV used in our experiments is lower than the band gap of ZnTe at room temperature (2.23eV). As a result, the displacive excitation, which requires real optical transitions resulting in an appreciable density of non-equilibrium carriers, is not effective for far below the band gap excitation.

In an attempt to find further evidence for the realization of TPBS in ZnTe we decided to use double pulse excitation for the mode under study. This is usually achieved by the coherent control technique, where the first pump pulse starts a coherent motion with all atoms moving in a synchronous manner, then the second pump pulse excites a phase-delayed replica of the first and finally the two interfere. Thus, by varying the phase difference (that is interpulse separation), it is possible to bring the superposition either to a destructive or a constructive interference and thus to control the final state attained after the double pulse excitation. To this end, we used two collinear pump pulses with a variable interpulse separation to modify TPBS. The change in the two-phonon amplitude as the interpulse separation was varied is shown in Fig. 3. As seen from this figure, the resulting coherent amplitude in ZnTe varies almost harmonically in a way that the TPBS can be either significantly enhanced or almost suppressed, which happens at integer and half-integer oscillatory periods, respectively. It is interesting that when the two-phonon bound oscillations are substantially suppressed, an obvious mode-beating appears in the time-domain signal (Fig. 4 (a)). As shown in Fourier (Fig. 4 (b)) and continuous wavelet transforms (Fig. 4(c)), there are two distinct modes with frequencies of 3.67 THz and 3.26 THz, respectively, at the interpulse separation of 680 fs (~2.5 $T_C$). Note that the frequency for the low frequency mode exactly coincides with that of the 2TA(X) overtone [9]. The appearance of the 2TA(X) overtone gives further evidence on the repulsive binding in the TPBS located at higher frequency, because the frequency obtained from the time-domain spectroscopy can be sometimes different from that from the conventional frequency-domain spectroscopy. Since both TPBS and unbound two-phonon states are optically active, the ratio between the both contributions depends on the strength of the anharmonicity $\Gamma$ relative to the dispersion $W$. In the present case, the amplitude of the TPBS is much higher than that of the unbound two-phonon states. As a result, the oscillation of the unbound two-phonon states is hidden under the strong the TPBS oscillation. Only when the TPBS oscillation is suppressed by the optical control, the oscillation of the unbound two-phonon states can be clearly observed.

## IV. CONCLUSION

In summary, we have demonstrated for the first time, in the model case of ZnTe(110), that the TPBS can be generated by an ultrashort laser pulse. The existence of the TPBS, consisting of two repulsively bound TA(X) phonons is supported by *i*) the symmetry of observed excitation, which has full symmetry of the crystal point group, and *ii*) the energy, which is larger than that of the 2TA(X) overtone due to a repulsive binding. Furthermore, by using double pump pulses with a variable delay, the bound state can be manipulated in such a way that both the bounded and unbounded two phonon states are observed when the TPBS is suppressed.

**FIGURE CAPTIONS**

FIG. 1 (color online). Experimental results from a single-pulse excitation demonstrating a typical transient transmission change in ZnTe(110) obtained using a pump power of 30 mW. To clarify the oscillatory part, the signal is 100 times zoomed in the upper inset. The lower inset shows the Fourier transformed spectrum of the oscillatory part.

FIG. 2 (color online). Dependence of the TPBS amplitude on the polarization angle (a) and pump power (b). The dashed line is just a guide to the eye. Note that the frequency (not shown) of the TPBS has no considerable change for different angles in (a) and different pump powers in (b).

FIG. 3 (color online). Modulation of the amplitude of coherent two-phonon bound oscillations as a function of the interpulse separation $\Delta t$ in units of the oscillation period $T_C$=272 fs.

FIG. 4 (color online). The oscillatory signal (a), its Fourier (b) and continuous wavelet (c) transforms for double pulse excitation at $\Delta t$=2.5 $T_C$, providing direct evidence of the coexistence of bound and unbound two-phonon states. In continuous wavelet transform the Morlet function with the wavenumber of 25 is adopted to resolve frequency.

# REFERENCES

* To whom all correspondence should be addressed: nakamura.k.ai@m.titech.ac.jp

[1] V.M. Agranovich and I.I. Lalov, Uspekhi Fiz.Nauk **146**, 267 (1985) [Sov.Phys.Uspekhi **28**, 484 (1985)].

[2] R. S. Krishnan, Proc. Ind. Acad. Sci. **24**, 25 (1946).

[3] M. H. Cohen and J. Ruvalds, Phys.Rev.Lett. **23**, 1378 (1969).

[4] K. L. Ngai, A. K. Ganguly, and J. Ruvalds, Phys.Rev. **B 10**, 3280 (1974).

[5] J. C. Kimball, C. Y. Fong, and Y. R. Shen, Phys.Rev. **B 23**, 4946 (1981).

[6] F. Bogani *et al*., Phys.Rev**. B 42**, 2307 (1990).

[7] M.L. Giernaert, G.M. Gale, and C. Flytzanis, Phys.Rev.Lett. **52**, 815 (1984).

[8] G. M. Gale, *et al*., Phys.Rev.Lett. **54**, 823 (1985).

[9] J. C. Irwin and J. LaCombe, J. Appl.Phys. **41**, 1444 (1970).

[10] R. L. Schmidt, B. D. McCombe, and M. Cardona, Phys.Rev. **B 11**, 746 (1975).

[11] R. Brazis and D. Nausewicz, Optical Materials **30**, 789 (2008).

[12] M. Schall, M. Walther, and P. Uhd Jepsen, Phys.Rev. **B 64**, 094301 (2001).

[13] Y. S. Lim, *et al*., Phys.Rev. **B 68**, 153308 (2003).

[14] N. Kamaraju, *et al*., J. Appl.Phys. **107**, 103102 (2010).

[15] H. Takahashi, et al., Solid State Commun. **149**, 1955 (2009).

[16] O. A. Dubovskii and A. V. Orlov, Fiz. Tverd. Tela. **36**, 3131 (1994).

[17] X. Hu and F. Nori, Physica **B 263-264**, 16 (1999).

[18] O. V. Misochko, J. Hu, and K. G. Nakamura, Phys. Lett. A **375** 4141 (2011)

[19] R. Merlin, Solid State Commun. **102**, 207 (1997).

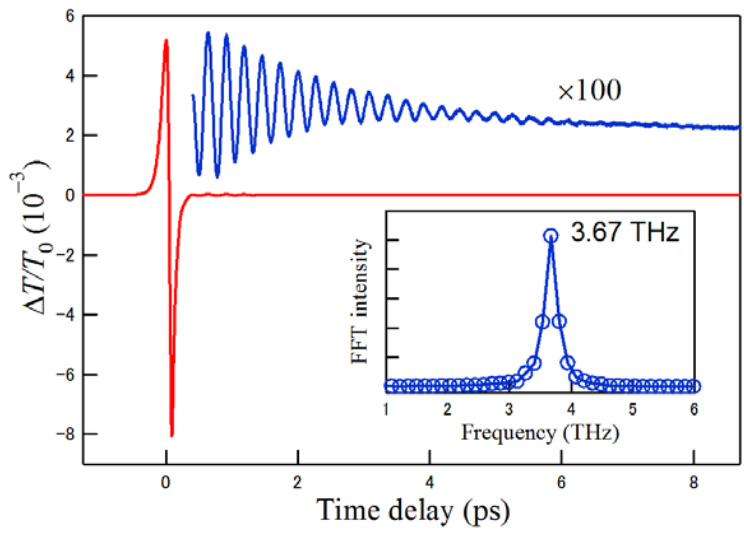

**Figure 1, Hu,** *et al.*

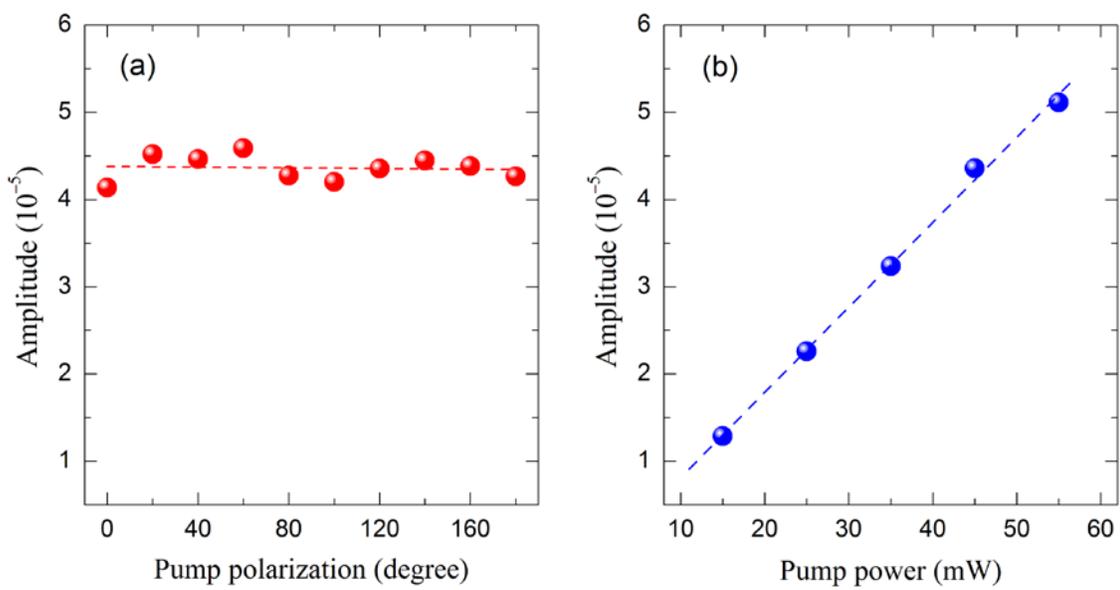

**Figure 2, Hu,** *et al.*

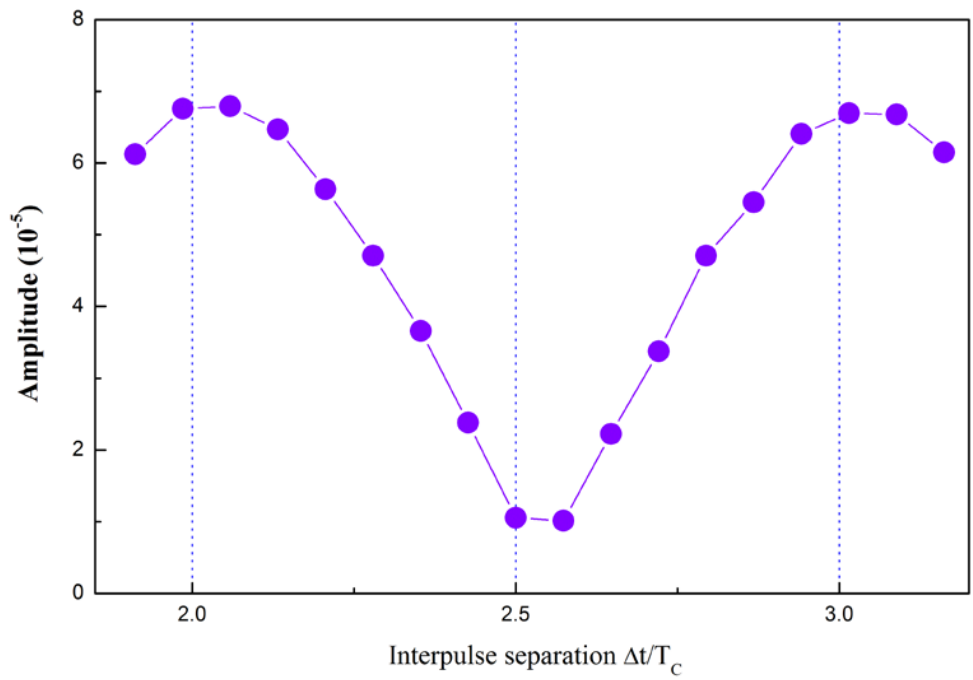

**Figure 3, Hu,** *et al.*

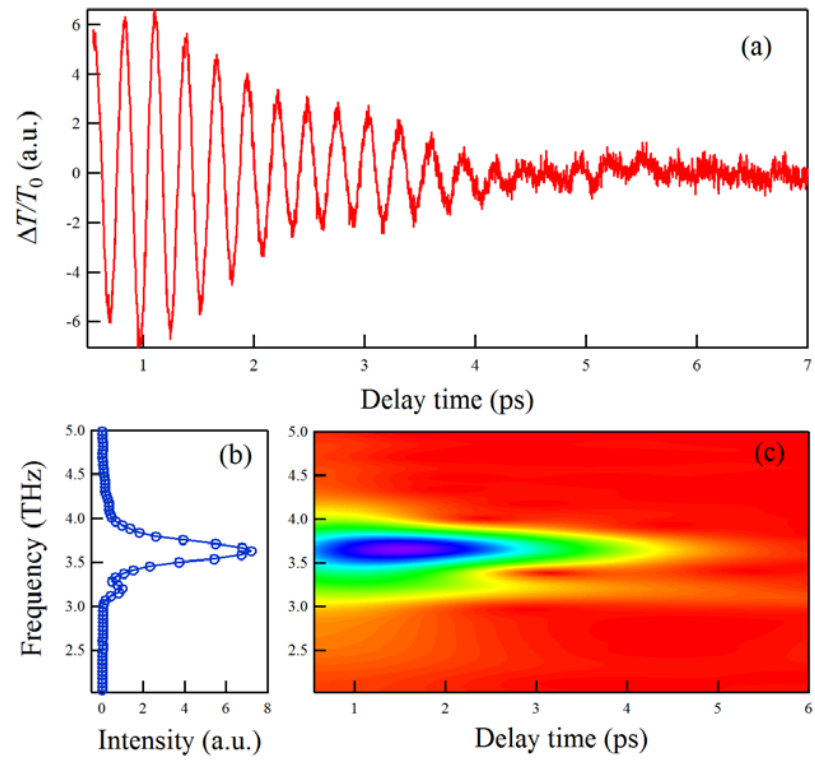

**Figure 4, Hu,** *et al.*